\documentclass[aps,pre,twocolumn,superscriptaddress,noshowpacs]{revtex4-1}
\usepackage{amsmath}
\usepackage{amssymb}
\usepackage{tikz}
\usepackage{graphicx}
\usepackage{dcolumn}
\usepackage{bm}
\usepackage{epsfig}
\usepackage{psfrag}
\usepackage{latexsym}
\def\II{\hbox{$1\hskip -1.2pt\vrule depth 0pt height 1.6ex width 0.7pt\vrule depth 0pt height 0.3pt width 0.12em$}}
\newcommand{\reffig}[1]{\mbox{Fig.~\ref{#1}}}
\newcommand{\refeq}[1]{\mbox{Eq.~(\ref{#1})}}
\newcommand{\refsec}[1]{\mbox{Sec.~\ref{#1}}}

\newcommand{\be}{\begin{equation}}
\newcommand{\ee}{\end{equation}}
\newcommand{\ba}{\begin{eqnarray}}
\newcommand{\ea}{\end{eqnarray}}
\renewcommand{\Re}{\mathrm{Re}}

\newcommand{\T}{${\mathcal T}$}
\newcommand{\Ti}{${\mathcal T\, }$}

\begin{document}
\draft
\title{\bf How $\mathcal{T}$-invariance violation leads to an enhanced backscattering with increasing openness of a wave-chaotic system}
\author{Ma{\l}gorzata Bia{\l}ous}
\affiliation{%
Institute of Physics, Polish Academy of Sciences, Aleja Lotnik\'{o}w 32/46, 02-668 Warszawa, Poland
}
\author{Barbara Dietz}
\altaffiliation{
email:  Dietz@lzu.edu.cn}
\affiliation{%
School of Physical Science and Technology, and Key
Laboratory for Magnetism and Magnetic Materials of MOE, Lanzhou University,
Lanzhou, Gansu 730000, China
}
\author{Leszek Sirko}
\altaffiliation{
email:  sirko@ifpan.edu.pl}
\affiliation{%
Institute of Physics, Polish Academy of Sciences, Aleja Lotnik\'{o}w 32/46, 02-668 Warszawa, Poland
}

\date{\today}

\bigskip

\begin{abstract}
We report on the experimental investigation of the dependence of the elastic enhancement, i.e., enhancement of scattering in backward direction over scattering in other directions of a wave-chaotic system with partially violated time-reversal (\T) invariance on its openness. The elastic enhancement factor is a characteristic of quantum chaotic scattering which is of particular importance in experiments, like compound-nuclear reactions, where only cross sections, i.e., the moduli of the associated scattering matrix elements are accessible. In the experiment a quantum billiard with the shape of a quarter bow-tie, which generates a chaotic dynamics, is emulated by a flat microwave cavity. Partial \T-invariance violation of varying strength $0\leq \xi\lesssim 1$ is induced by two magnetized ferrites. The openness is controlled by increasing the number $M$ of open channels, $2\leq M\leq9$, while keeping the internal absorption unchanged. We investigate the elastic enhancement as function of $\xi$ and find that for a fixed $M$ it decreases with increasing time-reversal invariance violation, whereas it increases with increasing openness beyond a certain value of $\xi\gtrsim 0.2$. The latter result is surprising because it is opposite to that observed in systems with preserved \Ti invariance ($\xi =0$). We come to the conclusion that the effect of \Ti-invariance violation on the elastic enhancement then dominates over the openness, which is crucial for experiments which rely on enhanced backscattering, since, generally, a decrease of the openness is unfeasible. Motivated by these experimental results we, furthermore, performed theoretical investigations based on random matrix theory which confirm our findings.      

\end{abstract}

\bigskip
\maketitle
\section{Introduction\label{Intro}}
The features of the classical dynamics of a closed Hamiltonian system are reflected in the spectral fluctuation properties of the corresponding quantum system~\cite{Berry1977,Casati1980,Bohigas1984}. For a chaotic dynamics they are predicted to coincide with those of random matrices from the Gaussian orthogonal ensemble (GOE) if the system is time-reversal (\T) invariant. This was confirmed in numerous experimental and numerical studies of nuclear systems~\cite{Brody1981,Bohigas1983,Gomez2011,Dietz2017}, and of various other systems~\cite{LesHouches1989,Guhr1998,Stoeckmann1999,Haake2010,Vina1998,Zimmermann1988,Hul2004,Frisch2014,Mur2015,Naubereit2018,Lawniczak2019}. We report on experiments with flat microwave resonators referred to as microwave billiards~\cite{Stoeckmann1990,Sridhar1991,Graef1992,So1995,Sirko1997} emulating quantum billiards~\cite{Sinai1970,Bunimovich1979,Berry1981}. Systems with violated time-reversal (\Ti) invariance are described by the Gaussian unitary ensemble (GUE), as observed, e.g., in atoms in a constant external field~\cite{Sacha1999}, in quantum dots~\cite{Marcus1992,Pluhar1995}, in Rydberg excitons~\cite{Assmann2016} in copper oxide crystals, nuclear reactions~\cite{French1985,Mitchell2010}, microwave networks~\cite{Allgaier2014,Bialous2016,Lawniczak2019a} and in microwave billiards~\cite{So1995,Stoffregen1995,Dietz2019}. A random matrix theory (RMT) description was also developed for partially violated \Ti invariance~\cite{Pandey1991,Lenz1992,Altland1993,Bohigas1995,Pluhar1995} and applied recently to experimental data obtained with a superconducting microwave billiard~\cite{Dietz2019}.

Similar observations concerning the descriptiveness by RMT were also made for quantum chaotic scattering systems. In fact, RMT was originally introduced in the field of nuclear physics~\cite{Guhr1998}. Nuclear-reaction experiments yield cross sections of which the fluctuations have been investigated thoroughly for the \T-invariant case and compared to RMT predictions for quantum scattering systems~\cite{Ericson1960,Lynn1968} and for other many-body systems~\cite{Stania2005,Madronero2005,Celardo2007}. The case of \T-invariance violation (TIV) was considered in Ref.~\cite{French1985} for nuclear spectra, in Refs.~\cite{Ericson1966,Mahaux1966,Witsch1967,Blanke1983,Boose1986} for compound-nuclear reactions, and in Refs.~\cite{Pluhar1995,Bergmann1984,Rosny2005} for other devices. Analytical expressions have been derived within RMT for the scattering ($S$)-matrix autocorrelation function for preserved~\cite{Verbaarschot1985} and partially violated \Ti invariance~\cite{Dietz2009a,Dietz2010} and verified experimentally with microwave billiards. This is possible because the scattering formalism describing them~\cite{Albeverio1996} coincides with that of compound nuclear reactions~\cite{Mahaux1969} and both the modulus and phase of $S$-matrix elements are accessible, whereas in compound nuclear reactions only cross sections, that is, the modulus, can be determined. Furthermore, large data sets may be obtained for systems with preserved, partially or completely violated \Ti invariance. The analogy has been used in numerous experiments~\cite{Blumel1990,Mendez2003,Schaefer2003,Kuhl2005,Hul2005,Lawniczak2008,Dietz2009a,Dietz2010,Bialous2016} for the investigation of statistical properties of the $S$ matrix using as indicator for TIV that the principles of reciprocity, $S_{ab}=S_{ba}$, and of detailed balance, $\vert S_{ab}\vert =\vert S_{ba}\vert$, no longer hold. 

Another statistical measure of quantum chaotic scattering, which is of particular importance, e.g., in nuclear physics because it can be determined from cross-section measurements and does not depend on the mean resonance spacing, is the elastic enhancement factor (EEF) as a measure for the enhancement of elastic scattering processes, that is, scattering in backward direction or back to the initial scattering channel over inelastic ones to other directions or scattering channels. Such an enhancement was observed in compound-nucleus cross sections~\cite{Satchler1963,Kretschmer1978} and, actually, is an universal wave phenomen~\cite{Jin1990,Barabanenkov1991,Baranger1970,Fazio2017,Jacucci2018,Bergmann1984}. The elastic enhancement factor was proposed as a tool to characterize a scattering process by Moldauer~\cite{Moldauer1964} and serves as a probe of quantum chaos in nuclear physics~\cite{Verbaarschot1985,Verbaarschot1986,Kharkov2013} and in other fields~\cite{Fyodorov2005,Savin2006,Dietz2010,Zhirov2015}. 

The nuclear cross section $\sigma_{ab}$ provides a measure for the probability of a nuclear-reaction process involving an incoming particle in scattering channel $b$ scattered, e.g., at a nucleus, thus forming a compound nucleus and eventually decaying into a residual nucleus and a particle in scattering channel $a$. Its energy average is expressed in the framework of the Hauser-Feshbach theory~\cite{Hauser1952,Hofmann1975} in terms of the $S$ matrix elements $S_{ab}(\nu;\eta,\gamma,\xi)) = \langle S_{ab} \rangle + S^{\rm fl}_{ab}(\nu;\eta,\gamma,\xi)$, $\sigma^{\rm fl}_{ab}=\sigma_{ab}-\langle\sigma_{ab}\rangle=\vert S^{\rm fl}_{ab}\vert^2\equiv C_{ab}(0;\eta,\gamma,\xi)$. Here, $\nu$ denotes the energy of the incoming particles in a nuclear reaction or the microwave frequency in a microwave billiard, and $C_{ab}(\varepsilon;\eta,\gamma,\xi)$ is the $S$-matrix autocorrelation function. Both $S_{ab}$ and $C_{ab}$ depend on the openness, that is, on the number $M$ of open channels and the strength of their coupling to the environment given in terms of the parameter $\eta$ and the absorption $\gamma$, and on the size of TIV quantified by a parameter $\xi$. The EEF can be expressed in terms of $C_{ab}(0;\eta,\gamma,\xi)$, 
\begin{equation}
        F_M(\eta,\gamma,\xi) = 
	\frac{\sqrt{C_{aa}(0;\eta,\gamma,\xi)\, C_{bb}(0;\eta,\gamma,\xi)}}{\sqrt{C_{ab}(0;\eta,\gamma,\xi)\, C_{ba}(0;\eta,\gamma,\xi)}}
        \label{defF}
\end{equation}
In the sequel we will suppress the dependence of $S_{ab}(\nu)=S_{ab}(\nu;\eta,\gamma,\xi)$ and $C_{ab}(\epsilon)=C_{ab}(\epsilon;\eta,\gamma,\xi)$ on $\eta,\gamma$ and $\xi$ like is commonly done. 

Analytical results are obtained for the EEF by inserting those for $C_{ab}(\epsilon)$~\cite{Verbaarschot1985,Dietz2009a} interpolating between preserved ($\beta=1,\, \xi=0$) and completely violated ($\beta =2,\, \xi\simeq 1$) \Ti invariance~\cite{Dietz2010}. The limiting values, attained for well isolated resonances, where the resonance width $\Gamma$ is small compared to the average resonance spacing $d$, and for strongly overlapping ones are known, 
\begin{equation}
        F^{(\beta)}_M(\eta,\gamma) \rightarrow \left\{
        \begin{array}{rl}
                1 + 2/\beta & \ {\rm for}\ \Gamma/d \ll 1 \\
                    2/\beta & \ {\rm for}\ \Gamma/d \gg 1
        \end{array}
        \right.\, .
        \label{Flimits}
\end{equation}
Accordingly, a value of the EEF below 2 indicates TIV~\cite{Dietz2010}. For the case of partial TIV the features of $F^{(\beta)}_M(\eta,\gamma,\xi)$ as function of $\xi$ \emph{and} $M$ are not yet well understood. The objective of the present article is to fill this gap by performing thorough experimental and RMT studies of $F_M(\eta,\gamma,\xi)$ in the $(\eta,\xi)$ plane.

Properties of the EEF have been investigated experimentally in microwave networks~\cite{Lawniczak2010}  and in microwave resonators~\cite{Zheng2006,Dietz2010,Yeh2013,Lawniczak2015,Bialous2019} with two attached antennas, that is, $M=2$ open channels, of similar size of the coupling to the interior states, which is quantified by the transmission coefficients $T_c=1-\vert\langle S_{cc}\rangle\vert^2\simeq T,\, c=1,2$. Weak coupling corresponds to $T\simeq 0$ and perfect coupling to $T=1$~\cite{Verbaarschot1985,Verbaarschot1986}. Recently, we investigated the EEF in a microwave billiard~\cite{Bialous2019} as function of the openness by varying $M$ and thus $\eta =MT$~\cite{Kharkov2013} while keeping the absorption fixed. In the present article we report on the first experimental study of the EEF for increasing $M$ in the presence of \Ti violation. Such experiments are of particular relevance for nuclear physics and, generally, experiments relying on enhanced backscattering, because there typically the number of open channels can be large. In~\refsec{Experiment} we introduce the experimental setup and then present experimental results in~\refsec{Results}. Then we explain how we determined the experimental parameters, i.e., the openness $\eta$, absorption $\gamma$ and size of TIV $\xi$ on the basis of analytical RMT results and then finally discuss the experimental and RMT results for the enhancement factor in the $(\eta,\xi)$ plane.  

\section{Experimental setup\label{Experiment}}
\begin{figure}[tb]
\includegraphics[width=0.7\linewidth]{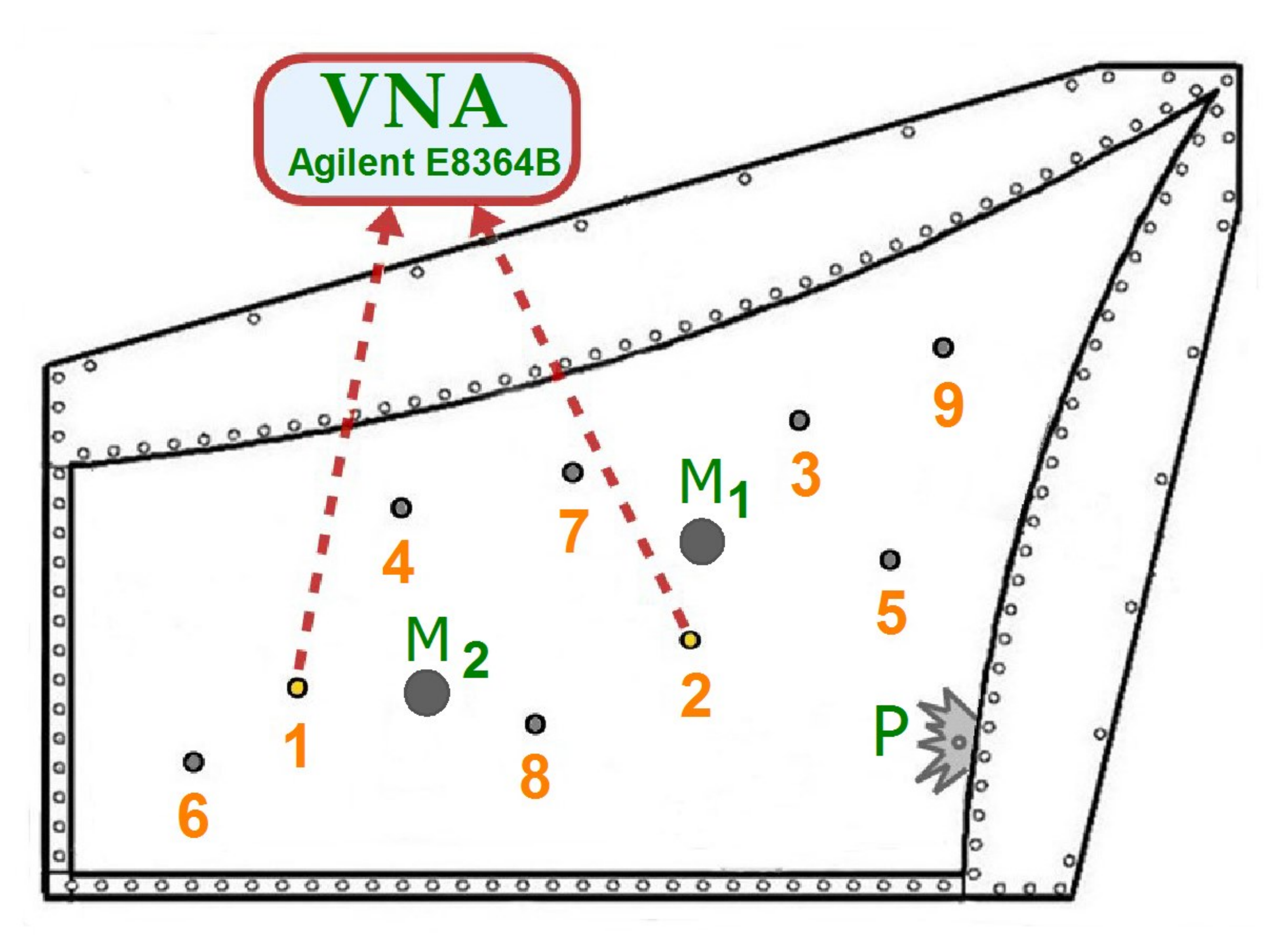}
	\caption{ Schematic top view of the flat microwave resonator with the shape of a quarter bow-tie billiard which has a chaotic classical dynamics. See~\refsec{Experiment} for a detailed description of the experimental setup.  
}\label{Fig1}
\end{figure}

We used the same microwave billiard as in our previous studies~\cite{Bialous2019}. A schematic top view of the cavity is shown in~\reffig{Fig1}. It has the shape of a quarter bow-tie billiard with area $\mathcal{A}=1828.5$~cm$^2$ and perimeter $\mathcal{L}=202.3$~cm of which the classical dynamics is fully chaotic. The height of the cavity is $h=1.2$~cm corresponding to a cut-off frequency of $\nu_{max}=c/2d\simeq 12.49$~GHz with $c$ the speed of light in vacuum. Below $\nu_{max}$ only transverse-magnetic modes are excited so that the Helmholtz equation describing the microwave billiard is scalar and mathematically identical to the Schr\"odinger equation of the quantum billiard of corresponding shape. The inner surface of the cavity is covered with a 20~$\mu$m layer of silver to reduce internal absorption. The top lid of the cavity has 9 randomly distributed holes of same size marked from 1 to 9 in~\reffig{Fig1}. The sub-unitary two-port $S$ matrix $S_{ab},\, a\ne b,\, a,b\in\{1,\dots , 9\}$ was measured yielding $C_{ab}(\epsilon)$ and the associated EEF. For this, wire antennas of length 5.8~mm and pin diameter 0.9~mm are attached to the holes $a,b$ and connected to an Agilent E8364B Vector Network Analyzer (VNA) with flexible microwave cables. The additional open channels are realized by successively attaching to the other holes according to their numbering antennas of the same size but shunted with 50~$\Omega$ loads. Since identical antennas are used, the associated transmission coefficients take similar values, so that $\eta =MT$. The amplitudes of the resonances in the spectra $\vert S_{ab}(\nu)\vert$ depend on the size of the electric field at the positions of the emitting and receiving antennas. Since the resonator has the shape of a chaotic billiard, and thus the average electric field intensity is distributed uniformly over the whole billiard area, the EEF does not depend on the choice of positions of the measuring antennas. Therefore, we will present results only for the measurements where we chose antenna positions at $a=1$ and $b=2$. 

\begin{figure}[tb]
\includegraphics[width=0.6\linewidth]{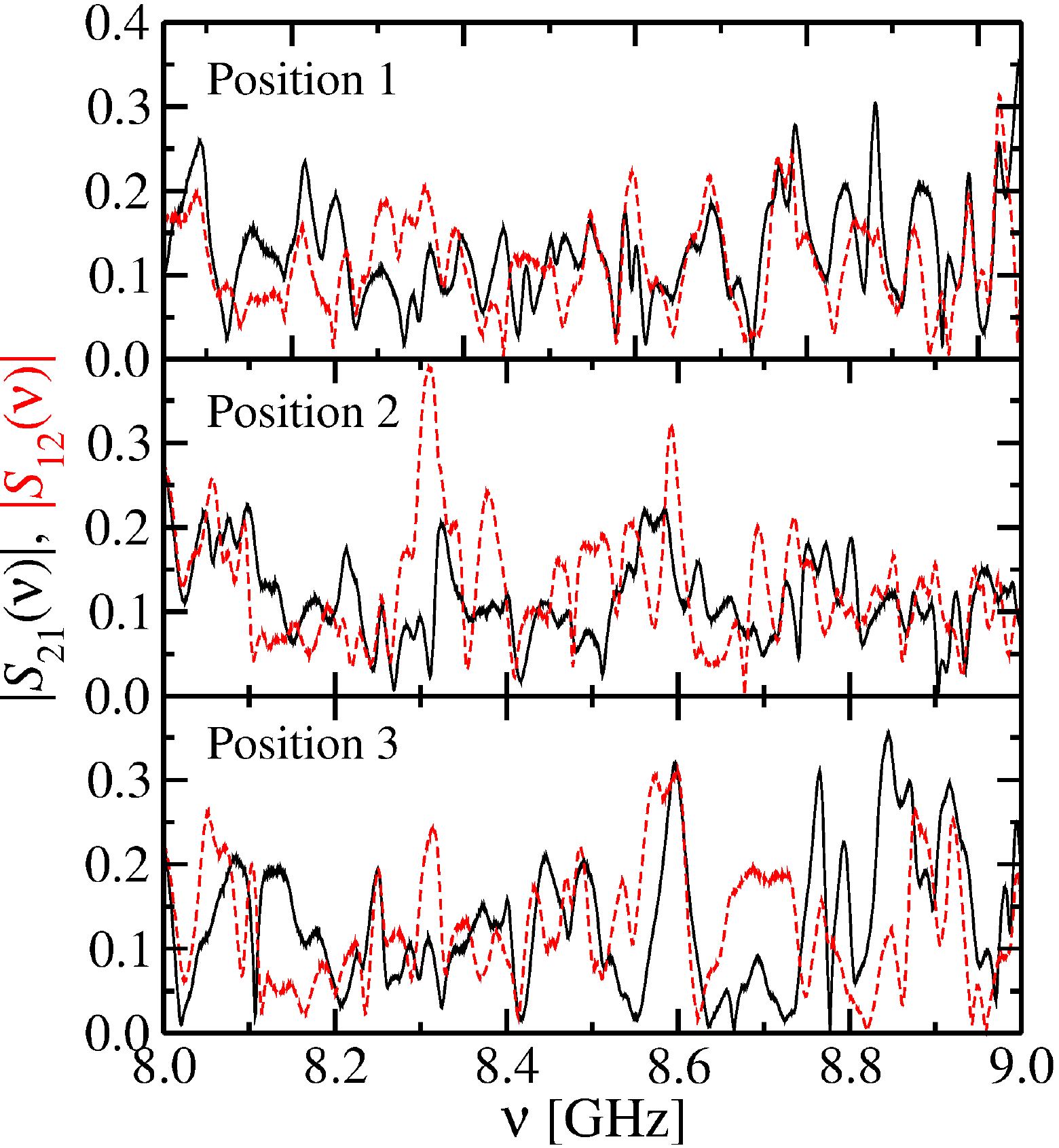}
	\caption{ Transmission spectra $\vert S_{21}(\nu)\vert$ (black full lines) and $\vert S_{12}(\nu)\vert$ (red dashed lines) for three consecutive positions of the perturber in the microwave frequency range $\nu\in [8,9]$~GHz. Violation of the principle of detailed balance and, thus of TIV is clearly visible.
}\label{Fig1a}
\end{figure}

All measurements are performed in the frequency range $\nu\in [6,12]$~GHz. To realize an ensemble of 100 microwave billiards of varying shape, a metallic perturber marked by a 'P' in~\reffig{Fig1} with area 9~cm$^2$, perimeter 21~cm is placed with its straight boundary part of length 2~cm at the sidewall inside the cavity and moved stepwise along the wall with an external magnet~\cite{Bialous2019}. The size of the steps of 2~cm is of the order of the wave length of the microwaves, which varies between 5~cm at 6~GHz and 2.5~cm at 12~GHz, and thus induces sufficiently large changes in the spectra, as illustrated in~\reffig{Fig1a}, to attain statistical independence of all realizations. In order to induce TIV, two cylindrical NiZn ferrites (manufactured by SAMWHA, South Korea) with diameter $33$~mm, height $6$~mm and saturation magnetization $2600$~Oe are inserted into the cavity and magnetized by an external homogeneous magnetic field of strength $B\simeq 495$~mT generated by a pair of NdFeB magnets of type N42 with coercity 11850~Oe placed above and below the cavity at the ferrite positions marked by M$_1$ and M$_2$ in~\reffig{Fig1}. Here, the positions of the ferrites were chosen such that largest possible TIV is achieved. The magnetic field $B$ induces a macroscopic magnetization in the ferrites which precesses around $B$ with the Larmor frequency $\omega_{o}=\gamma_G B$, where $\gamma_G\simeq g_{\rm eff}\cdot 14$~GHz/T and $g_{\rm eff}\simeq 2.3$  denote the gyromagnetic ratio and the Lande factor, respectively, thereby causing the appearance of a ferromagnetic resonance at $\nu_{\rm FR}\approx 15.9$~GHz. The closer the microwave frequency is to it the stronger is the size of TIV. However, as clearly visible in~\reffig{Fig1a} showing $S_{21}(\nu)$ and $S_{12}(\nu)$ in the frequency range $\nu\in[8,9]$~GHz, the principle of detailed balance does not hold already well below $\nu_{\rm FR}$.

\section{Experimental results\label{Results}}
\begin{figure}[tb]
{\includegraphics[width=\linewidth]{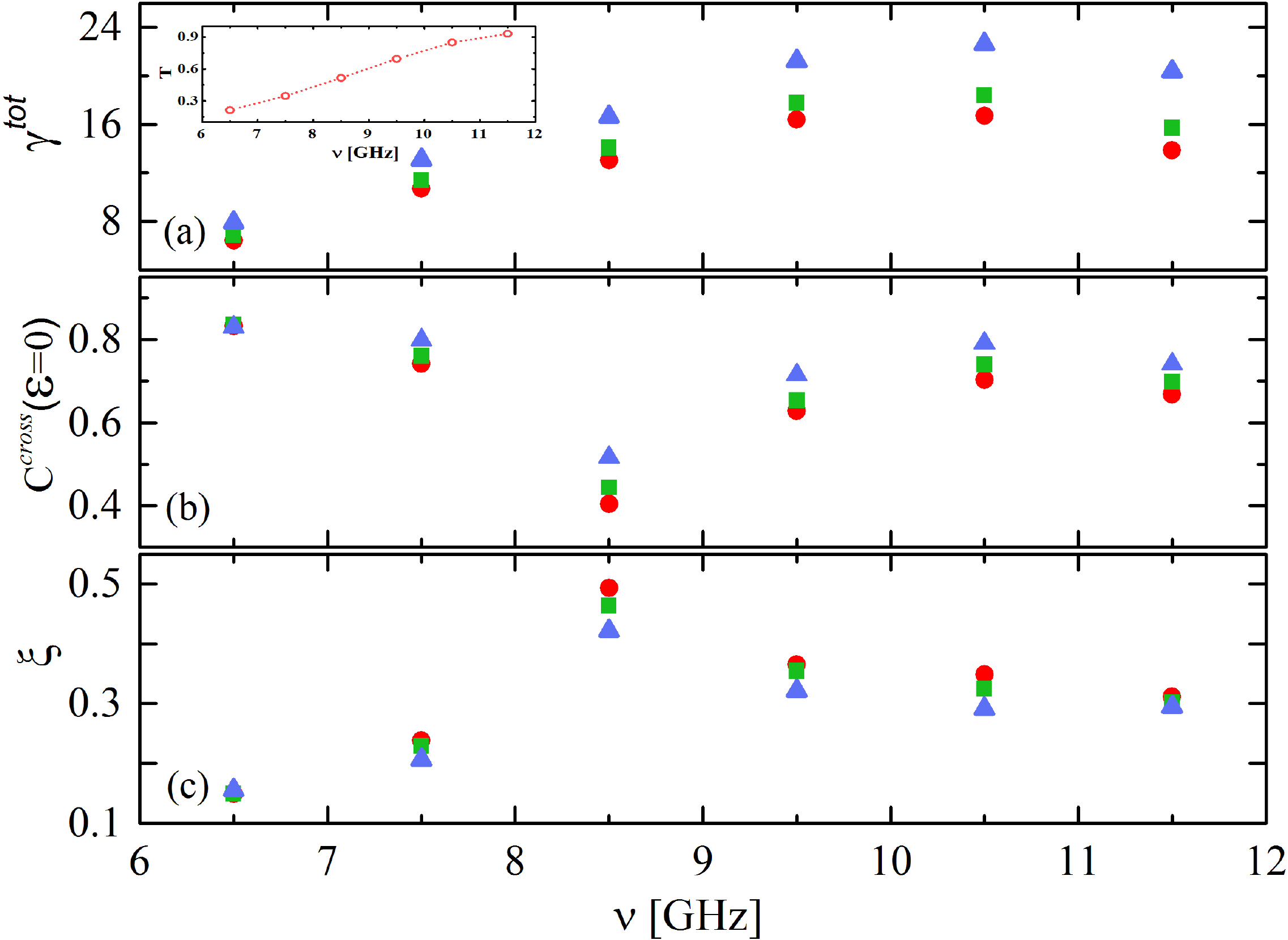}}
\caption{(a) Rescaled resonance width $\gamma^{tot}$ versus the microwave frequency $\nu$. The inset shows the average transmission coefficients. (b) Experimentally determined cross-correlation coefficient $C^{cross}(0)$ for $M=2$ (red circles), 4 (green squares) and 9 (blue triangles) open channels. (c) Same as (b) for the strength $\xi$ of TIV.
}
\label{Fig3}
\end{figure}

We used the cross-correlation coefficient $C^{cross}_{12}(0)=C^{cross}_{12}(\varepsilon=0;\eta,\gamma,\xi)$,
\begin{equation}
\label{Eq.3}
C^{cross}_{12}(0)=\frac{\Re[{\langle S^{\rm fl}_{12}(\nu)\, S^{\rm fl*}_{21}(\nu)\rangle]}}{\sqrt{\langle|(S^{\rm fl}_{12}(\nu)|^2\rangle\langle|(S^{\rm fl}_{21}(\nu)|^2\rangle}},
\end{equation}
as a measure for the size of TIV. It equals unity for \T-invariant systems, and approaches zero with increasing size of TIV. We verified that the experimental cross-correlation coefficient, average resonance width and transmission coefficients are approximately constant in a frequency range of 1~GHz and accordingly evaluated the average of $C^{cross}_{12}(0)$ over the 100 cavity realizations in 1~GHz windows. The result is shown in~\reffig{Fig3}~(b). It exhibits a broad minimum in the frequency range $\nu\in[8,9]$~GHz implying that strongest TIV is induced by the magnetized ferrites at about half the value of $\nu_{\rm FR}$. This may be attributed to the occurrence of modes trapped inside the ferrite~\cite{Dietz2019}.
Figure~\ref{Fig3} (a) shows the rescaled resonance widths $\gamma^{tot}=\frac{2\pi}{d}\Gamma$. It results from two contributions, namely the width $\Gamma_a$ due to absorption of the electromagnetic waves in the walls of the cavity, ferrites and the metallic perturber and the escape width $\Gamma_{\rm esc}$ originating from the additional open channels describing the coupling of the internal modes to the continuum. Absorption is accounted for by $\Lambda\gg 1$ weakly open, identical fictitious channels with transmission coefficients $T_f\ll T$~\cite{Verbaarschot1986,Savin2006}. Note, that choosing three different values for $T_f$ to account for the absorption properties of the cavity walls, the ferrites and the perturber which are made from different materials, where the fractions of fictitious channels are given by those of their perimeters~\cite{Schaefer2003}, yields similar results. The absorption strength is related to $\Gamma_a$ according to the Weisskopf relation via $\gamma=\frac{2\pi\Gamma_a}{d}=\Lambda T_f$~\cite{Blatt1952}. The openness $\eta =MT$~\cite{Kharkov2013} may be expressed in terms of the Heisenberg time $t_H=\frac{2\pi}{d}$ and the dwell time $t_W=\frac{1}{\Gamma_{\rm esc}}$ which gives the time an incoming microwave spends inside the cavity before it escapes through one of the $M$ open channels~\cite{Savin2006}, $\eta=t_H/t_W$. In terms of the  Weisskopf formula the escape width is given by $\frac{2\pi}{d}\Gamma_{\rm esc}=MT$, so that $\gamma^{tot}=MT+\Lambda T_f\equiv \eta+\gamma$. 

\begin{figure}[tb]
\begin{center}
\rotatebox{0}{\includegraphics[width=\linewidth]{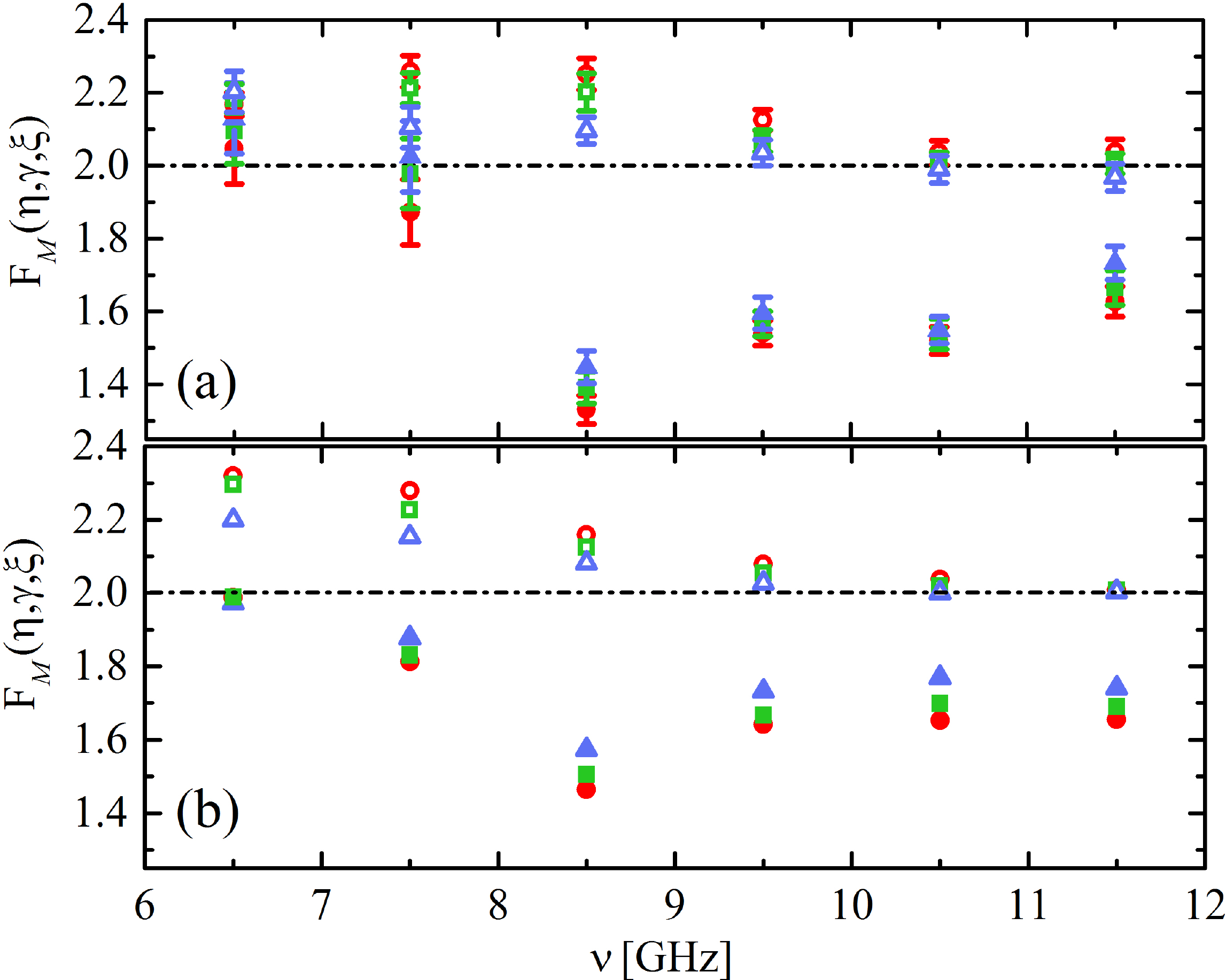}}
\caption{
(a) Elastic enhancement factor deduced from the two-port scattering matrix $\hat{S}$ measured in 1~GHz windows for $M=2$ (red circles), $M=4$ (green squares), and $M=9$ (blue triangles) open channels, respectively. Empty and full symbols were obtained without~\cite{Bialous2019} and with a magnetized ferrite inside the cavity. Each point is obtained by averaging over the 100 microwave billiard realizations. The error bars indicate the standard deviations. The black dash-dotted line separates the cases of preserved and violated \T-invariance; see~\refeq{Flimits}. (b) Same as (a) for the RMT results.
}\label{Fig4}
\end{center}
\end{figure}

The experimental EEF $F_M(\eta,\gamma,\xi)$ is obtained by averaging over the ensemble of 100 different cavity realizations in 1~GHz windows. The result is shown in~\reffig{Fig4}~(a) for $M=2$ (red circles), $M=4$ (green squares) and $M=9$ (blue triangles) open channels. Here, the empty and full symbols show the results for experiments without and with magnetized ferrite, respectively, and the error bars indicate the standard deviation. The black dash-dotted line separates the cases of preserved and violated \T-invariance. The value of $F_M(\eta,\gamma,\xi)$ is below two above 6~GHz and it exhibits a pronounced minimum in the frequency interval [8,9]~GHz. Furthermore, while for the \T-invariant case the value of the enhancement decreases with increasing $M$ as expected from~\refeq{Flimits}, surprisingly the opposite behavior is observed for the case of partial TIV. In order to confirm this behavior and for a better understanding of its origin and of the occurrence of the minimum we performed studies based on RMT. 

\section{Random matrix theory approach\label{RMT}}
The input parameters of the RMT model are the transmission coefficients $T=T_{a}\simeq T_{b}$ associated with antennas $a$ and $b$, which are determined from the reflection spectra, $T_c\simeq T$ of the remaining $M-2$ open channels, the absorption $\gamma=\Lambda T_{f}$ and the \T-violation parameter $\xi$. The sizes of $\gamma$ and $\xi$ are determined by comparing the distribution of the experimental reflection coefficients $S_{11},\, S_{22}$ and the cross-correlation coefficient to analytical and numerical RMT results~\cite{Fyodorov2005,Dietz2009a,Dietz2010,Kumar2013}. Note, that in Ref.~\cite{Dietz2010} the absorption strength $\gamma$ was determined from the resonance widths. This is not possible for the experiments presented in this article because it is too large ($6\lesssim\gamma\lesssim 15$) due to the presence of the ferrites that consist of lossy material leading to a considerable degradation of the quality factor, especially in the vicinity of a ferromagnetic resonance and of trapped modes. 

The RMT results were obtained based on the $S$-matrix approach~\cite{Mahaux1969} which was developed in the context of compound nuclear reactions and also applies to microwave resonators~\cite{Albeverio1996}, 
\be
\hat S(\nu) = \II - 2\pi i\hat W^\dagger(\nu\II+i\pi\hat W\hat W^\dagger-\hat H)^{-1}\hat W,
\ee
where $\hat S$ is $(M+\Lambda)$ dimensional and $\hat H$ denotes the $N$-dimensional Hamiltonian describing the closed microwave billiard. We present results for the properties of the sub-unitary $S$-matrix with entries $S_{ab},\,  a,b=1,2$. Quantum systems with a chaotic classical dynamics and partial TIV are described by an ensemble of $N\times N$-dimensional random matrices composed of real, symmetric and anti-symmetric random matrices $\hat H^{(S)}$ and $\hat H^{(A)}$~\cite{Dietz2009a}, respectively, 
\begin{equation}
        H_{\rm\mu\nu}=H_{\rm\mu\nu}^{(S)}+i\frac{\pi\xi}{\sqrt{N}}
        H_{\rm\mu\nu}^{(A)},
\label{eqn:hamiltonian}
\end{equation}
interpolating between GOE for $\xi =0$ and GUE for $\pi \xi / \sqrt{N} = 1$, where GUE is attained already for $\xi\simeq 1$~\cite{Dietz2010}. Furthermore, $\hat W$ accounts for the coupling of the $N$ resonator modes to their environment through the $M$ open and $\Lambda$ fictitious channels~\cite{Verbaarschot1986,Savin2006}. It is a $(M+\Lambda)\times N$ dimensional matrix with real and Gaussian distributed entries $W_{e\mu}$ of which the sum $\sum_{\mu=1}^NW_{e\mu}W_{e\mu}= Nv_e^2,\, e=1,\dots,M+\Lambda$ yields the transmission coefficients $T_e=\frac{4\pi^2v^2_{e}/d}{(1+\pi^2v^2_{e}/d)^2}$~\cite{Dietz2010}.

\begin{figure}[tb]
{\includegraphics[width=0.6\linewidth]{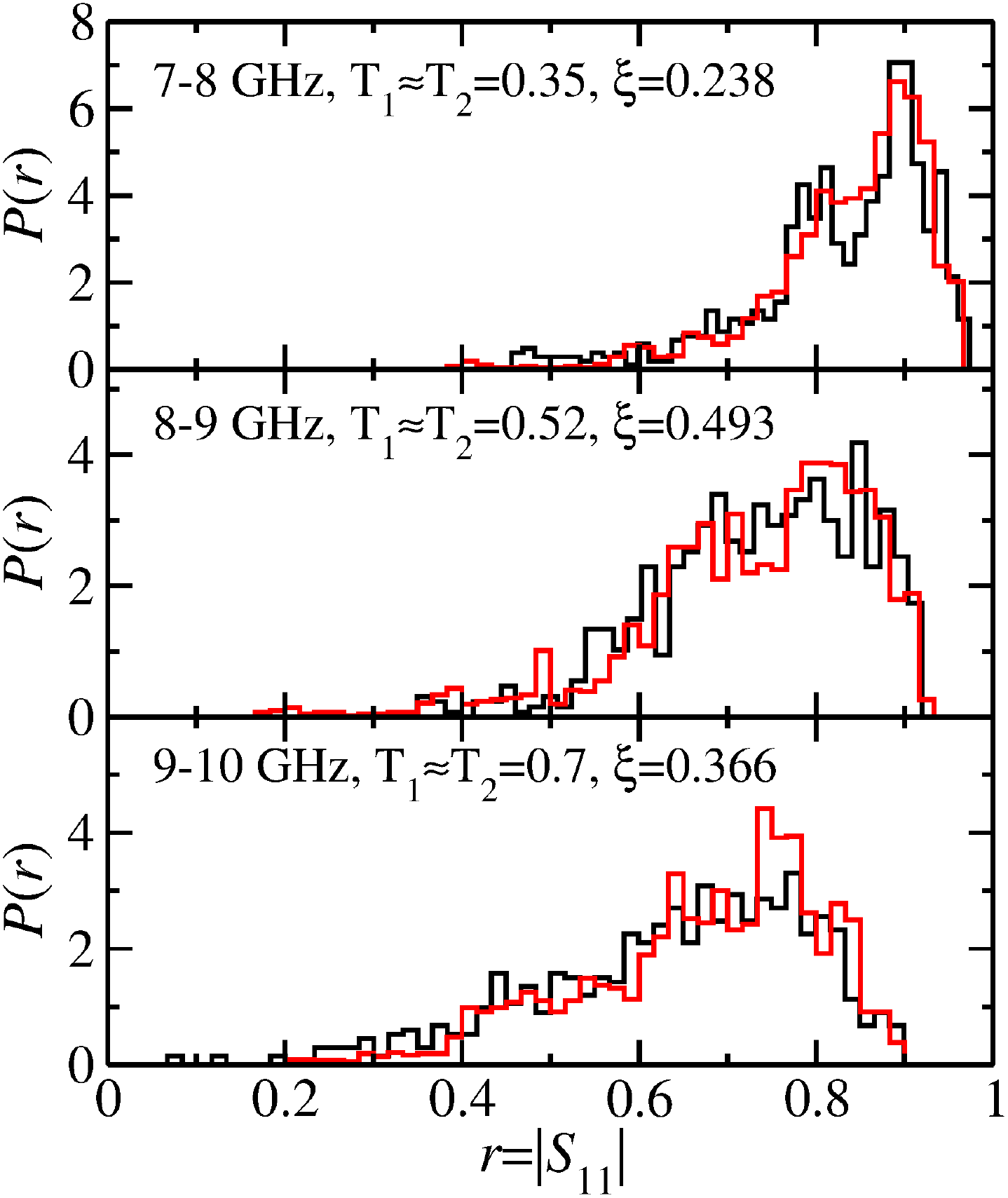}}
\caption{Experimental distributions of the modulus of $S_{11}$ for $M=2$ (black histograms) and the corresponding RMT results (red histograms) for different frequency ranges. The parameter values are given in the panels.
}
\label{Fig3a}
\end{figure}
\begin{figure}[tb]
\includegraphics[width=0.9\linewidth]{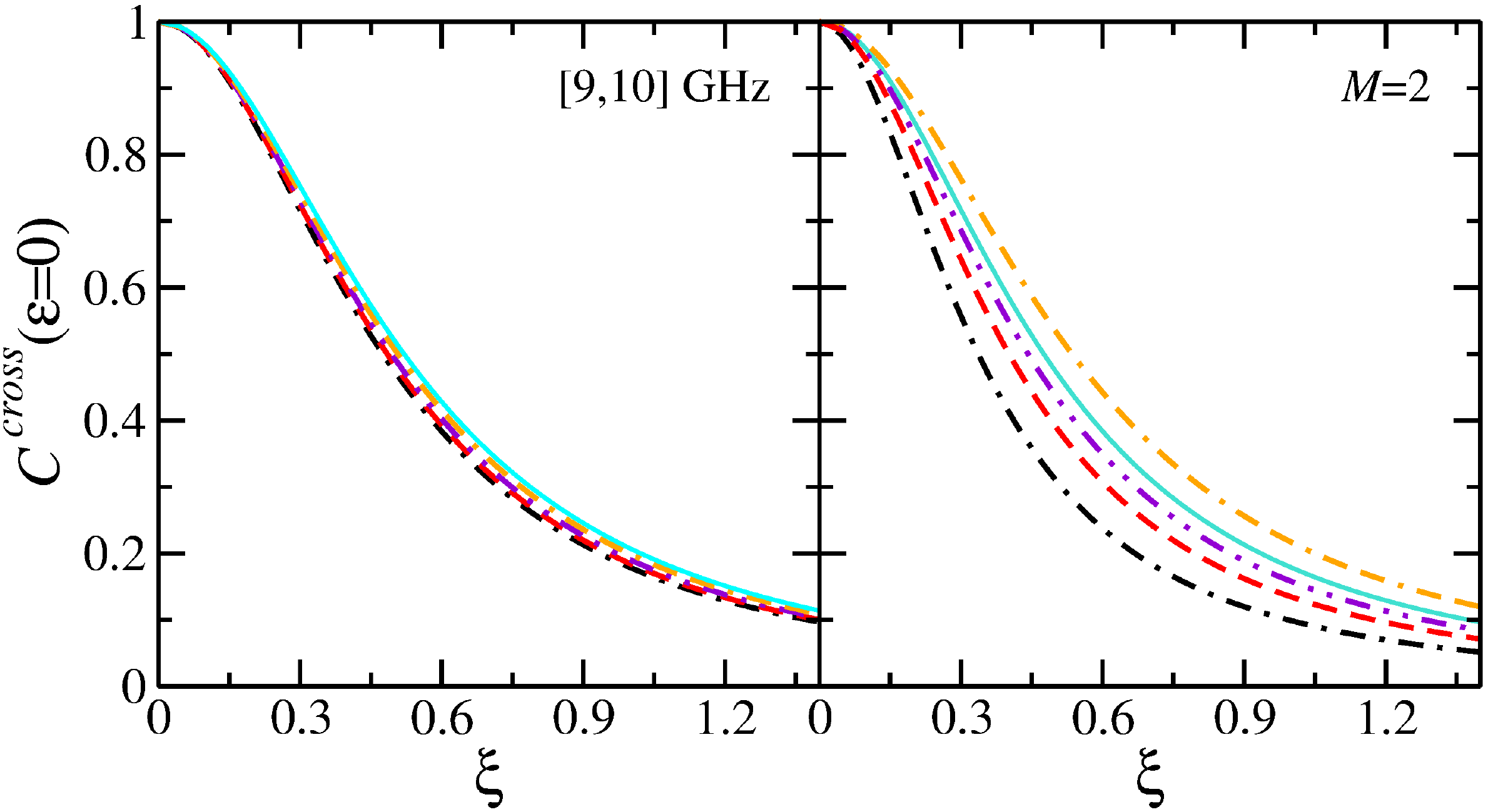}
\caption{Examples for the cross-correlation coefficients obtained from the analytical result~\cite{Dietz2009a}, using the experimental values of $\eta$ and $\gamma$ in a given frequency range $\nu\in[9,10]$~GHz [left: $M=2$ (black dash-dot line), $M=6$ (red dashed line), $M=9$ (orange dash-dash-dot line) and $M=10$ (cyan full line)] and for $M=2$ [right: $\nu=6.5$~GHz (black dash-dot line), $\nu=7.5$~GHz (red dashed line), $\nu=9.5$~GHz (violet dash-dot-dot line), $\nu=10.5$~GHz (orange dash-dash-dot line) and $\nu=11.5$~GHz (cyan full line)]. 
}
\label{Fig5a}
\end{figure}

Figure~\ref{Fig3a} shows the experimental distributions of the modulus of $S_{11}$ (black histogram) for a few examples. The red histograms show the RMT distributions best fitting the experimental ones. Figure~\ref{Fig3} (a) shows the resulting rescaled resonance widths $\gamma_{tot}=\gamma+\eta$ which indeed is considerably larger than in the experiments~\cite{Bialous2019} without ferrite. The largest absorption is $\gamma\simeq 15$ corresponding to strongest overlap of the resonances. Yet, the shape of the distributions of $\vert S_{11}\vert$ in~\reffig{Fig3} shows that the limit of Ericson fluctuations, where a bivariate Gaussian distribution is expected~\cite{Agassi1975}, is not yet reached. The experimental cross-correlation coefficients are shown in~\reffig{Fig3} (b) and (c) exhibits the corresponding values for the TIV parameter $\xi$. These were determined by proceeding as in Ref.~\cite{Dietz2010}, that is, we computed for each parameter set $(\eta,\gamma)$ the cross-correlation coefficient as function of $\xi$ using the analytical result of Ref.~\cite{Dietz2009a} and compared it with the experimental ones to determine $\xi$ as function of frequency. The left panel of~\reffig{Fig5a} shows the analytically determined cross-correlation coefficients in the frequency range$\nu\in[9,10]$~GHz, the right one for $M=2$ open channels, where $\eta$ and $\gamma$ were chosen as in the experiments. Both reflect the features exhibited by $\xi$ in~\reffig{Fig3}~(c) in view of~\reffig{Fig3}~(b).

The cross-correlation parameter $\xi$ has a pronounced peak in the frequency interval [8,9]~GHz. There, the strength of TIV is largest, $\xi\simeq 0.49$. Above this interval its value still is comparatively large, $\xi\simeq 0.35$. In a given frequency range the size of $\xi$ decreases with increasing $M$, that is, with openness $\eta$. Note, that the size of TIV induced by the magnetized ferrite depends on the coupling of the spins in the ferrite precessing about the external magnetic field to the rf magnetic-field components of the microwaves, which in turn depends on the electric field intensity in the vicinity of the ferrite~\cite{Dietz2019}, and is largest at a ferromagnetic resonance or when microwave modes are trapped inside the ferrite and between the ferrite and top plate. Increasing the number of open channels leads to an increasing loss of microwave power and thus to a decrease of the electric-field intensity which explains the degression  of $\xi$. To compute the EEF $F_M(\eta,\gamma,\xi)$ we used~\refeq{defF}, that is, inserted the thereby determined values for $\gamma$, $\xi$ and $\eta$ into the analytical result for the autocorrelation function~\cite{Dietz2009a}. The results are shown in~\reffig{Fig4}~(b). Empty symbols were obtained by setting $\xi =0$. The RMT results for $\xi\ne 0$ clearly reproduce the course of the experimental ones for $F_M(\eta,\gamma,\xi)$. The pronounced dip exhibited by $F_M(\eta,\gamma,\xi)$ in the frequency range $\nu\in[8,9]$~GHz coincides with that of largest achieved TIV, $\xi\simeq 0.49$. 
\begin{figure}[tb]
\begin{center}
\rotatebox{0}{\includegraphics[width=0.8\linewidth]{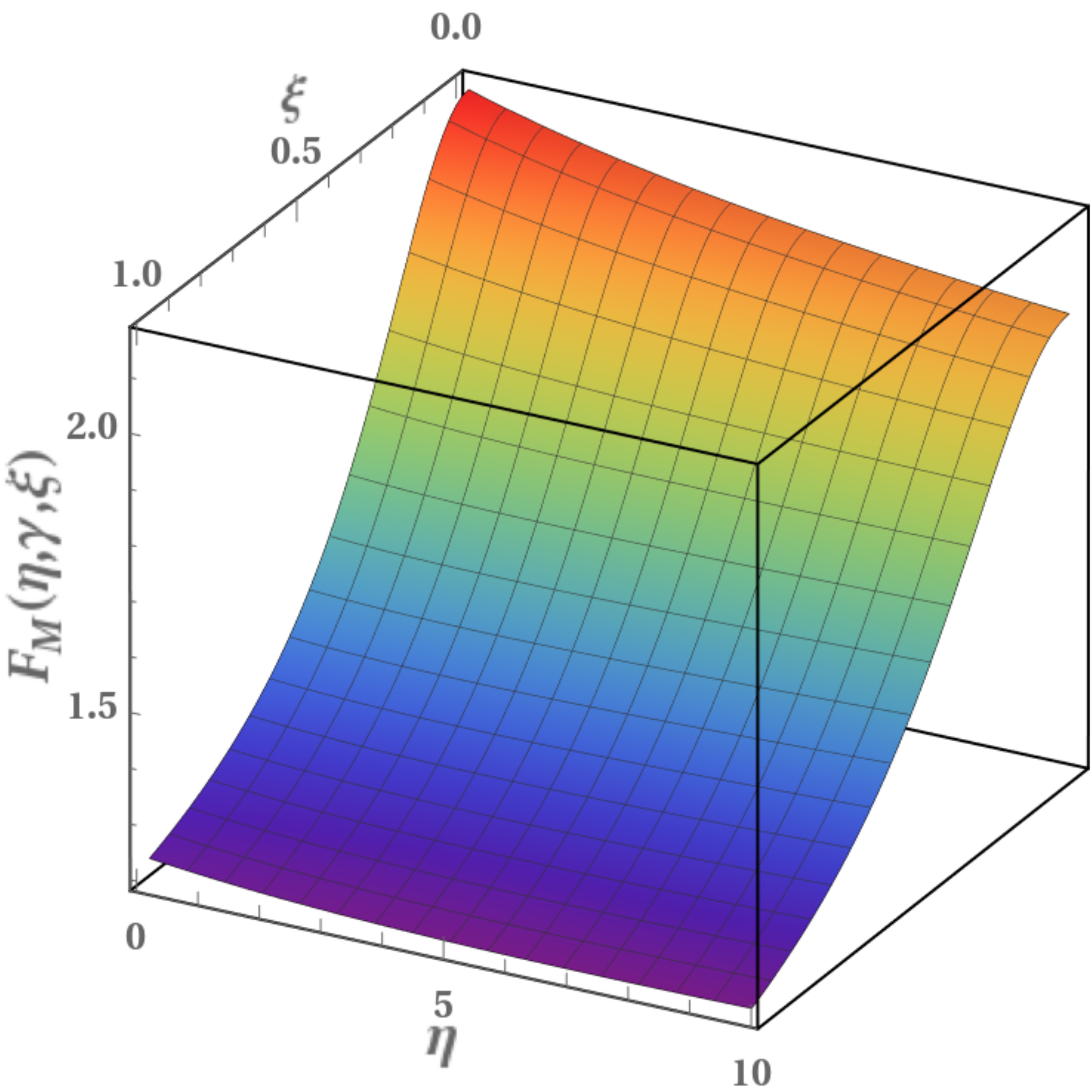}}
\caption{
Three-dimensional plot of the computed EEF $F_M(\eta,\gamma,\xi)$ for $M=10$ open channels versus $\eta= MT$ with $T\in [0,1]$ and $\xi\in [0,1]$. The total number of fictitious channels $\Lambda =50$ and absorption $\gamma =10$ were kept fixed. 
}\label{Fig6}
\end{center}
\end{figure}
\begin{figure}[tb]
\begin{center}
\rotatebox{0}{\includegraphics[width=0.8\linewidth]{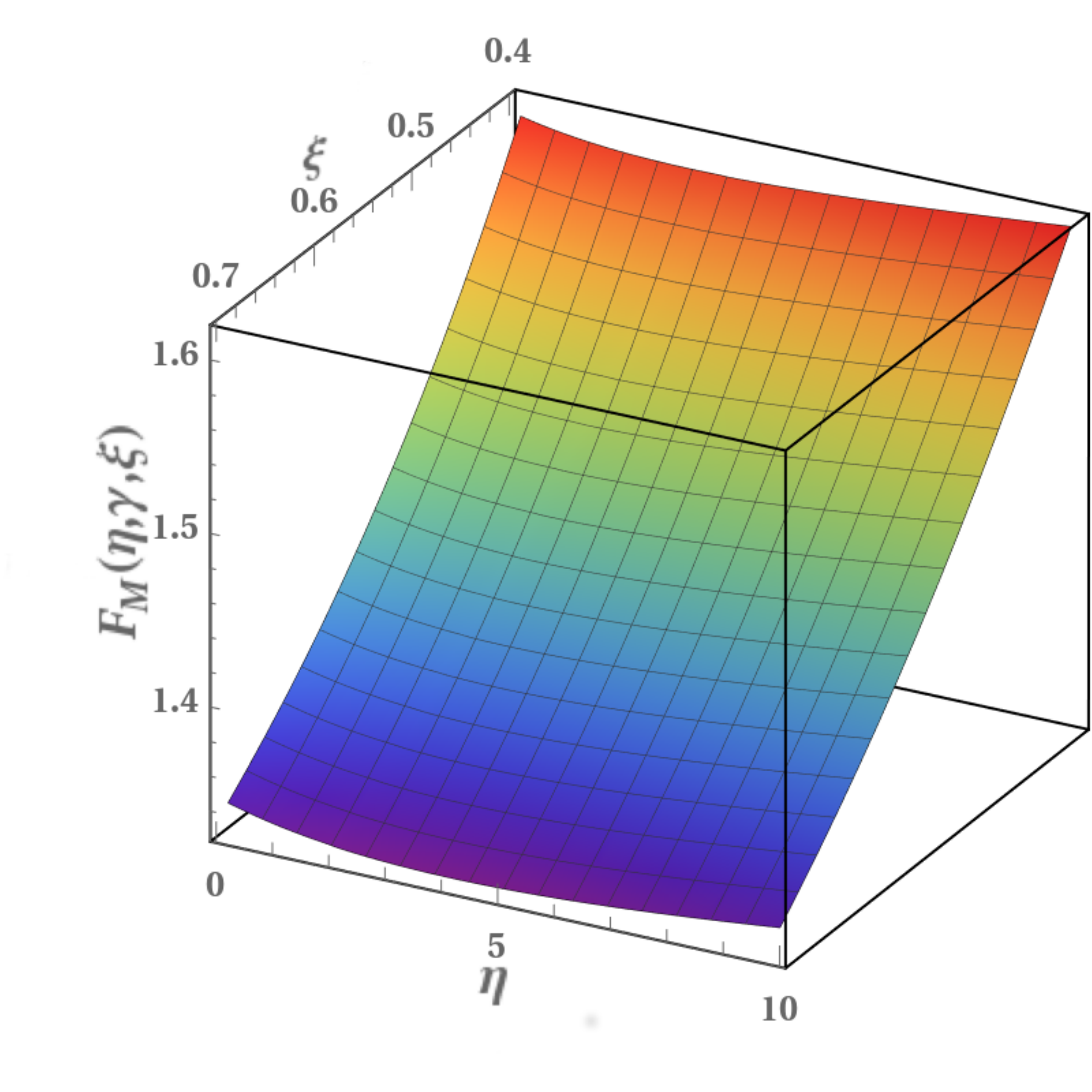}}
\caption{
Zoom of~\reffig{Fig6} into the region $0.4\leq\xi\leq 0.7$ of the computed EEF $F_M(\eta,\gamma,\xi)$.
}\label{Fig6a}
\end{center}
\end{figure}
\begin{figure}[tb]
\begin{center}
\rotatebox{0}{\includegraphics[height=0.6\linewidth]{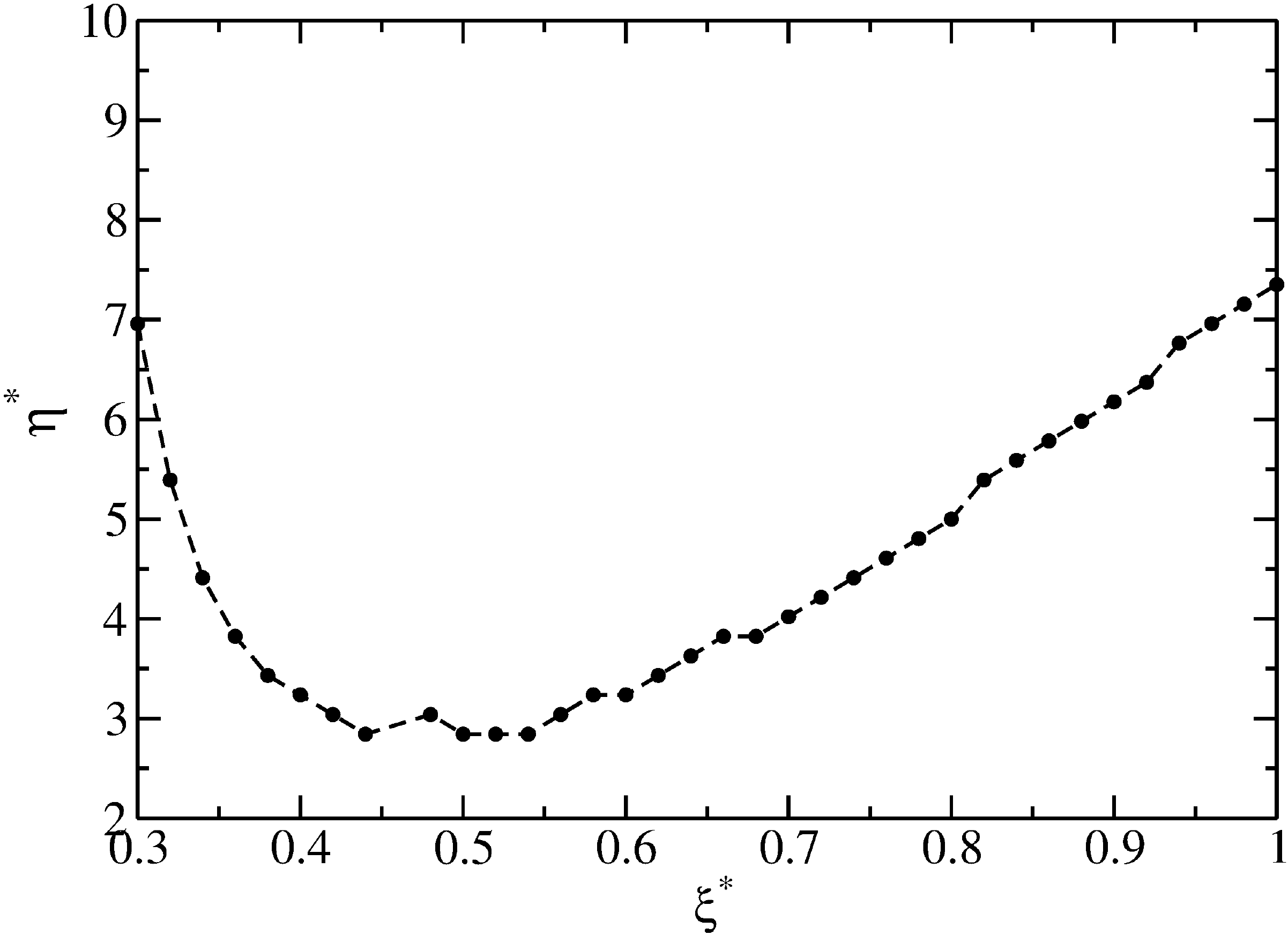}}
\caption{
Values of $(\eta,\xi)$ 	of the minimum exhibited for $\xi\geq 0.3$ by the enhancement factor $F_M(\eta,\gamma,\xi)$.
}\label{Fig6b}
\end{center}
\end{figure}
\section{Discussion and Conclusions\label{Concl}}
We investigate the EEF $F_M(\eta,\gamma,\xi)$ for a fully chaotic quarter bow-tie microwave billiard with partial TIV. It is induced by two magnetized ferrites and largest in the frequency range $\nu\in[8,9]$~GHz which is below the frequency of the ferromagnetic resonance at $\nu_{\rm FR}=15.9$~GHz, thus indicating that there the microwaves form a resonance inside and above the ferrite~\cite{Dietz2019}. For a fixed number $M$ of open channels the EEF decreases with increasing size of TIV, thus confirming the results for $M=2$ of Ref.~\cite{Dietz2010}. However, in distinction to the case of preserved \Ti invariance it increases for fixed $M$ with increasing absorption $\gamma=\gamma^{tot}-MT$ as clearly visible, e.g., in the frequency range $\nu\in[9,12]$~GHz where $\xi$ is approximately constant, and in a given frequency window with increasing $M$ for $\xi\gtrsim 0.2$. These observations are confirmed by RMT calculations. Figure~\ref{Fig6} exhibits the resulting $F_M(\eta,\gamma,\xi)$ in the $(\eta,\xi)$ parameter plane and~\reffig{Fig6a} a zoom into the region of $0.4\leq\xi\leq 0.7$. Here, $M=10$ and $\gamma=10$ were kept fixed while $T$ and $\xi$ were varied. These computations show that for a fixed value of $\eta$ $F_M(\eta,\gamma,\xi)$ indeed decreases with increasing $\xi$. Furthermore, it behaves like in the \Ti invariant case for $\xi\lesssim 0.3$ and then opposite, that is, for a given $\xi\geq 0.3$ it decreases with increasing $\eta$ until it reaches a minimum and then increases with $\eta$. The positions of the minima $(\xi^\ast,\eta^\ast)$ of $F_M(\eta,\gamma,\xi)$ in the $(\eta,\xi)$ plane are shown in~\reffig{Fig6b}. These results are in accordance with our experimental findings. Indeed, the experimental values of $\eta$ are larger than $\eta^\ast$ for $\xi > 0.2$. We may conclude, that the effect of TIV on the EEF dominates over that of the openness already for moderate values of $\xi$ and demonstrate thereby that the lower bound of $1\leq F_M(\eta,\gamma,\xi)\leq 2$ is not necessarily reached by increasing $\gamma^{tot}=\frac{2\pi\Gamma}{d}$ as might be expected from the behavior of the EEF for preserved \Ti invariance, but by increasing the size of TIV. Our findings are of relevance for experiments relying on enhanced backscattering in a multitude of open quantum and disordered systems with violated TIV, as they allow its increase by modifying the size of TIV instead of diminishing the openness which typically is impossible.   

Acknowledgments. This work was supported in part by the National Science Center, Poland, Grant No. UMO-2018/30/Q/ST2/00324. B.D. thanks the National Natural Science Foundation of China for financial support through Grants Nos. 11775100 and 11961131009.

\bibliography{TRIV_14_09.bbl}

\end{document}